\documentstyle[12pt]{article}

\title{Models for the evolution of free granular surfaces}

\author{R. Mulet\thanks{Permanent address:
 Superconductivity Laboratory, Physics Faculty-IMRE,
University of Havana,
10400 La Habana, Cuba} \hspace{0.1cm} 
and H. Herrmann\\
ICA1, University of Stuttgart,\\
Pfaffenwaldring 27, 70596 Stuttgart,\\
Germany}

\date{\today}

\begin{document}
\maketitle

\begin{abstract}
We introduce two sets of continuum equations to describe
granular flow on a free surface and study their
properties. The equations derived from a microscopic picture 
that includes  jumps and a mobility threshold, account for 
ripple and crater formation.

{\em Keywords: surface flow. granular materials}
\end{abstract}

\section{Introduction}
One of the apparently simple problems for which there is no consensus yet
on which are the correct continuum equations of motion is the flux
of grains on a granular surface (see for example ref. \cite{BCRE,Cargese,Boutreaux}). This flux can be driven by a fluid,
like air, by gravity or by an initial impact. It modifies the granular
surface itself either by deposition or by erosion. We will consider in this
paper a gravity driven flow with variable initial energy and consider 
deposition only. In addition we want to be in the limit of small flux.
One realization is the slow buildup of a pile from a point source.

A discrete model was proposed in ref.
\cite{Alonso_97} in which particles jump
down stairs and deposit if their energy is below a certain material
dependent threshold. From this model a set of continuum equations
was derived in ref.\cite{Grasselli_98} and applied to the problem of
impact craters. This continuum description, however, made an assumption
on the time dependence of the slope that only contained the threshold
in an indirect way, namely by inserting the known steady state
result. Also all the jumps were of length unity even if their
kinetic energy was different. In this paper we want to deal with these
defects of the equations in ref. \cite{Grasselli_98}
 by inserting the two mentioned
effects independently into the equations and investigating the
consequences analytically and numerically.

\section{Modelling jumps of grains}

We will introduce an extension to the model proposed in ref.
\cite{Grasselli_98}. There, particles of mass $m$ are added at the top of
a pile, and they move, under the action of gravity $g$, 
until their kinetic energy $e$ falls below a certain
threshold $U$ (related to the friction between grains) 
where they then stop. At each collision the kinetic energy decreases
by a factor proportional to the restitution coefficient of the
material $r$. Mathematically, it reads:

\begin{equation}
e(x+ \delta)=(e(x)+mg(h(x)-h(x+\delta)))r \label{eq:enerA_cla}
\end{equation}

It was assumed in ref. \cite{Grasselli_98} that the traveled distance
after each particle jump is constant. Here, trying to model a more
realistic situation, we assume that this distance
is proportional to the kinetic energy of the particle. 
So, we propose to modify equation (\ref{eq:enerA_cla}) to:

\begin{equation}
e(x+ qe)=(e(x)+mg(h(x)-h(x+qe)))r \label{eq:enerA_mod}
\end{equation}

\noindent where $q$ is a proportionality coefficient (in units of
$1\over mg$). Then,
writing eq. (\ref{eq:enerA_mod}) in differential form we obtain:

\begin{equation}
\frac{de}{dx}=\frac{1}{q}(r-1)+r \gamma(x) \label{eq:enerA_diff}
\end{equation}

\noindent where $\gamma(x)=-mgdh/dx$.

To this equation one adds in ref.\cite{Grasselli_98} an
evolution equation for the slope of the form:

\begin{equation}
\frac{d \gamma}{dx}=\Gamma (U-e(x)) \label{eq:angA_diff}
\end{equation}

\noindent where $\Gamma$ controls the rate of relaxation of the slope.
Boundary conditions were imposed trying to reproduce
real experimental situations of heap formation:
 $e(0)=e_o$ a value which is proportional to
the height from which the grains fall on the pile, and $\gamma(0)=0$.

The equations (\ref{eq:enerA_diff}) and (\ref{eq:angA_diff}),
constitute a pair of linear differential equations with constant 
coefficients. From them, the angle of a pile can be easily calculated:

\begin{equation}
\gamma(x)= \frac{1-r}{rq}(1-\cos(\sqrt{\Gamma r}x))-
                \sqrt{\Gamma r} (e_o-U)\sin(\sqrt{\Gamma r}x) 
                                \label{eq:angA_sol}
\end{equation}

\noindent and then, integrating equation (\ref{eq:angA_sol}) the shape of
the pile is:

\begin{equation}
h(x)=h(0) -\frac{1-r}{rqmg}x + \frac{1-r}{rqmg} 
                \frac{1}{\sqrt{\Gamma r}}\sin(\sqrt{\Gamma r}x)-
                \frac{1}{rmg}(e_o-U)\cos(\sqrt{\Gamma r}x)) 
                                \label{eq:altA_sol}
\end{equation}

This solution has two remarkable properties as shown in Figure 1. 
First of all  the
oscillatory behaviour superimposed to the usual linear decay of the
pile. These oscillations arise from the existence of a length scale $\Gamma$
which controles the rate of the relaxation of the slope while independently the particles jump a distance ($qe$) determined only by their kinetic energy. 
This phenomenon is similar to the formation of ripples due to
saltation \cite{Nishimori_97}, and it could be the first theoretical
prediction of ``gravity'' induced ripples.

Another important feature of eq. (6) is that it can reproduce a crater formation on the top of pile. In fact, in the limit $\sqrt{\Gamma r}x<<1$,
equation (\ref{eq:altA_sol}) transforms into:

\begin{equation}
mgh(x)=mgh(0)-\frac{1}{r}(e_o-U)(1-\frac{\Gamma r}{2} x^2) 
                                \label{eq:altA_sol_lim}
\end{equation}

\noindent showing the existence of a parabolic-like crater whose depth linearly increases with the initial kinetic energy of the particles.

\section{Modelling the mobility threshold}

In this model, we add to the equation for the energies
from ref.  \cite{Grasselli_98}

\begin{equation}
\frac{de}{dx}=\frac{r-1}{\delta}e+r\gamma(x)
        \label{eq:enerB_diff}
\end{equation}

\noindent  a new equation for the angles based on the assumption that the 
equilibrium local slope of the pile is constant, i.e. 

\begin{equation}
\phi(x+\delta)=\phi(x)
        \label{eq:angB_phi}
\end{equation}

\noindent where $\phi$ is the slope of the pile.

The evolution of the slope proceeds in the following way: if a
particle arrives at $x$, it has two possibilities (as described for a
similar model in ref. \cite{Alonso_97}), if $e>U$, it moves to $x+\delta$, then:

\begin{equation}
\phi(x+\delta)=\phi(x)-1
        \label{eq:angB_phi_d}
\end{equation}

\noindent or, it stops when $e<U$ and

\begin{equation}
\phi(x+\delta)=\phi(x)+1
        \label{eq:angB_phi_c}
\end{equation}

Equations (\ref{eq:angB_phi_d}) and (\ref{eq:angB_phi_c}) may be written in the following differential form:

\begin{equation}
\frac{d \phi}{dx}=\frac{1}{\delta} sgn(U-e)
        \label{eq:angB_phi_diff}
\end{equation}

We next solve the coupled differential equations 
(\ref{eq:enerB_diff}) and (\ref{eq:angB_phi_diff}) numerically. 
It is more convenient to write both equations in the same 
form. So dividing  eq. (\ref{eq:enerB_diff}) by $mg$, and noting 
that $\gamma= mg \phi$, we obtain the following equations:

\begin{equation}
\frac{dz}{dx}= \frac{r-1}{\delta}z +\frac{r}{\delta} \phi
        \label{eq:enerB_ns}
\end{equation}

\begin{equation}
\frac{d \phi}{dx}=\frac{1}{\delta} sgn(z_u -z)
\label{eq:phiB_ns}
\end{equation}

\noindent where $mgz=e$ and $mgz_u=U$.

Then, once $\phi(x)$ is known the height of the pile can be calculated
through integration of $\phi$ and obtained numerically.

To numerically 
solve the equations (\ref{eq:enerB_ns}) and (\ref{eq:phiB_ns}) the
function $sgn(x)$ was replaced by  the continous function
$tanh(\alpha x)$. Profiles calculated using different values of
$\alpha$ are shown in Figure 2. From the Figure we conclude that if
$\alpha>10^2$ all the numerical results are equivalent. To  be on the
secure side we used in the rest of the paper $\alpha=10^5$.

Figure 3 represents typical sandpile profiles obtained for different
values of the parameters. The bottom curve (where the crater is better
defined) represents a system where the grains have higher initial energy.

In figure 4 is plotted the equilibrium angle of the pile $\phi_{eq}$
as a function of
$U(1-r)/r$ for different values of $e_o$. As previously calculated in ref.
\cite{Alonso_97}, the angle of the pile is equal to $U(1-r)/r$ showing
the equivalence between our approach and
that proposed in \cite{Grasselli_98}.

The depth $\Delta h$ and the width $\Delta x$ of the craters, should only 
depend on the three parameters involved in the model $r,e_o$ and $U$. 
In fact, in figures 5 and 6,  
$\Delta h$ and  $\Delta x$ 
obtained using different sets of parameters $(r,e_o,U)$ collapse on 
the same curve following the scaling relations:

\begin{equation}
\Delta h = (\frac{e_o}{U}-1)^\alpha f((\frac{e_o}{U}-1)^{-\nu} r)
\label{eq:scaling_deltah}
\end{equation}

\noindent and

\begin{equation}
\Delta x = (\frac{e_o}{U}-1)^\beta g((\frac{e_o}{U}-1)^{-\nu} r)
\label{eq:scaling_deltax}
\end{equation}

\noindent where $\beta=2.0$, $\alpha=1.0$, $\nu=4.0$ and $f(x)$ 
and $g(x)$ are
scaling functions with the following properties $f(x) \sim 1 $ if $x<<1$ and 
$f(x) \sim 0$ if $x>>1$, idem for $g$. 
This indicates that around $e_0=U$ and $r=0$ we have 
critical behaviour with simple integer exponents. 

Physically equations (\ref{eq:scaling_deltah}) and (\ref{eq:scaling_deltax})
imply that in granular systems with negligible restitution coefficent ($r \sim 0$) the depth of the crater increases linearly
with $e_o/U-1$ and with $1/r$, in good agreement with
 eq.\ref{eq:altA_sol_lim} while the width of the craters 
increases parabolically ($\beta=2.0$). Therefore 
systems with small $e_o$, i.e. ($e_o \approx U$) or 
$r \approx 1$ do not develop a crater.

\section{Conclusions}
We have presented two sets of equations of motion for the limit
of dilute granular surface flow. The first set included a jump length 
proportional to the energy of the grains. This gave ripple
formation on the critical slope, a phenomenon which has not yet 
been observed experimentally. The second set was a cleaner way to
include the deposition threshold than done in ref. \cite{Grasselli_98}.
The results confirm that the previously used equation 
(introduced in ref. \cite{Grasselli_98}) 
gave qualitatively the right answer for the shape of the pile
which means that the picture of a characteristic length
for relaxation to the equilibrium angle is correct.

\section*{Acknowledgments}
We thank SFB 381 for financial support.

\newpage

\section*{Figure Captions}

\hspace{0.5cm} {\bf Figure 1} Graphical representation of equation
6. We chose $\frac{1-r}{rq}=0.1$,$\sqrt{\Gamma r}=1$ and $\frac{e_o-U}{r}=0.1$.
 
{\bf Figure 2} Shape of the pile calculated by integrating the solution
$\phi$ of equations (13) and (14) for $\alpha=1$ (upper curve) and
$\alpha=10$ and $100$ (bottom curve).

{\bf Figure 3} Shape of the pile for $r=0.03,z_U=0.05$, and (from top
to bottom) $z=0.05,0.10$ and $15$. The profiles are shifted in the $z$ axis
to distinguish them better.

{\bf Figure 4} Equilibrium angle of the pile $\phi_{eq}$ as a function
of $U(1-r)/r$ for different values of $e_o$,$r$ and $U$.

{\bf Figure 5} Data collapse of the depth of the craters using eq: 
(\ref{eq:scaling_deltah}).Parameters: ($r,e_o,U$) = 
$\triangle$ (0.01, 0.15, 0.05); 
$\Box$ (0.01, 0.25, 0.05); 
$\odot$ (0.01, 0.25, 0.10); 
$\bullet$ (0.01, 0.25, 0.15); 
$\times$ (0.03, 0.20, 0.10); 
$+$ (0.03, 0.25, 0.15); 
$\Box$ (0.05, 0.30, 0.10).

{\bf Figure 6} Data collapse of the depth of the crater using eq: (\ref{eq:scaling_deltax}). 
Parameters: ($r,e_o,U$) = 
$\triangle$ (0.01, 0.15, 0.05); 
$\Box$ (0.01, 0.25, 0.05); 
$\odot$ (0.01, 0.25, 0.10); 
$\bullet$ (0.01, 0.25, 0.15); 
$\times$ (0.03, 0.20, 0.10); 
$+$ (0.03, 0.25, 0.15); 
$\Box$ (0.05, 0.30, 0.10).

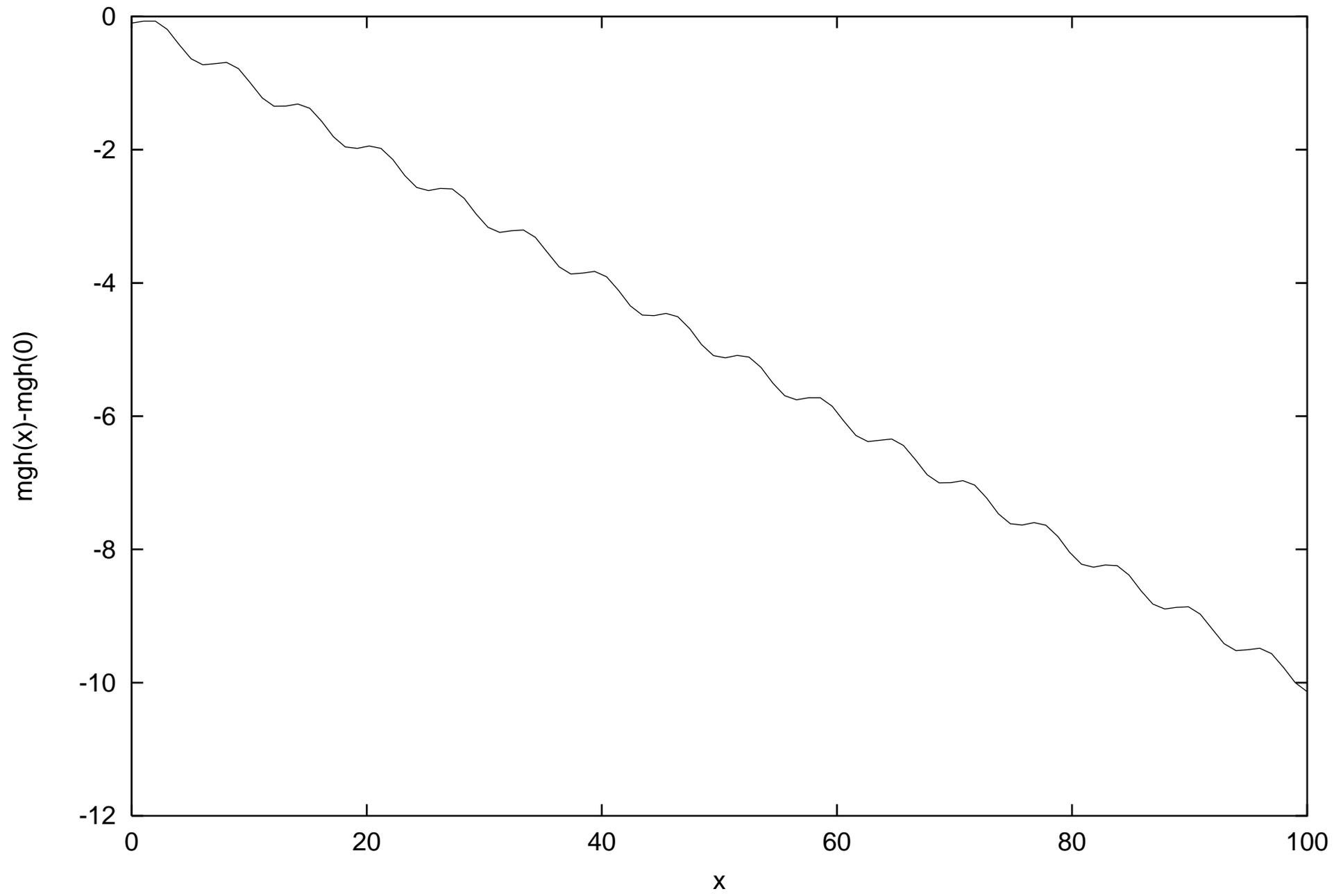
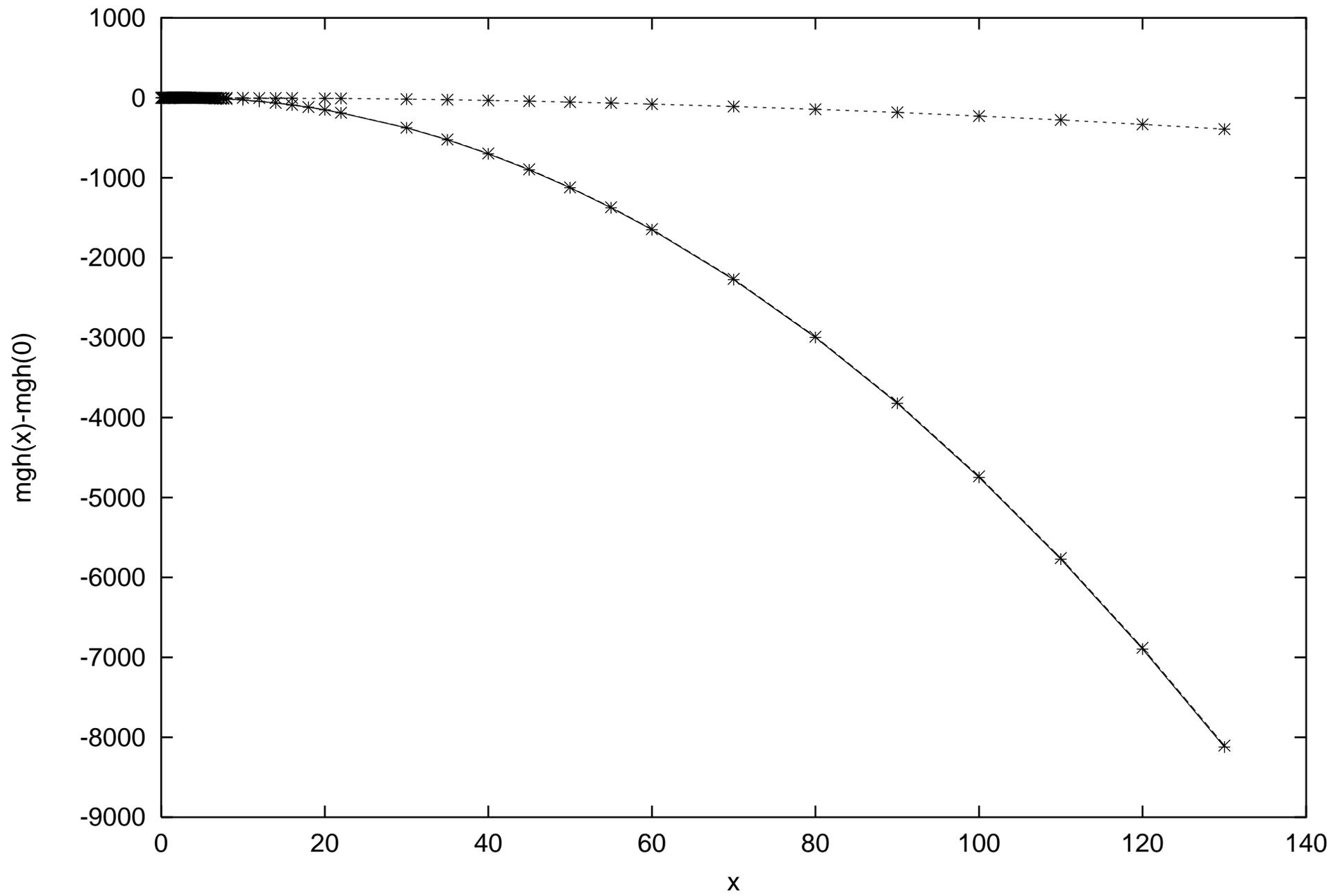
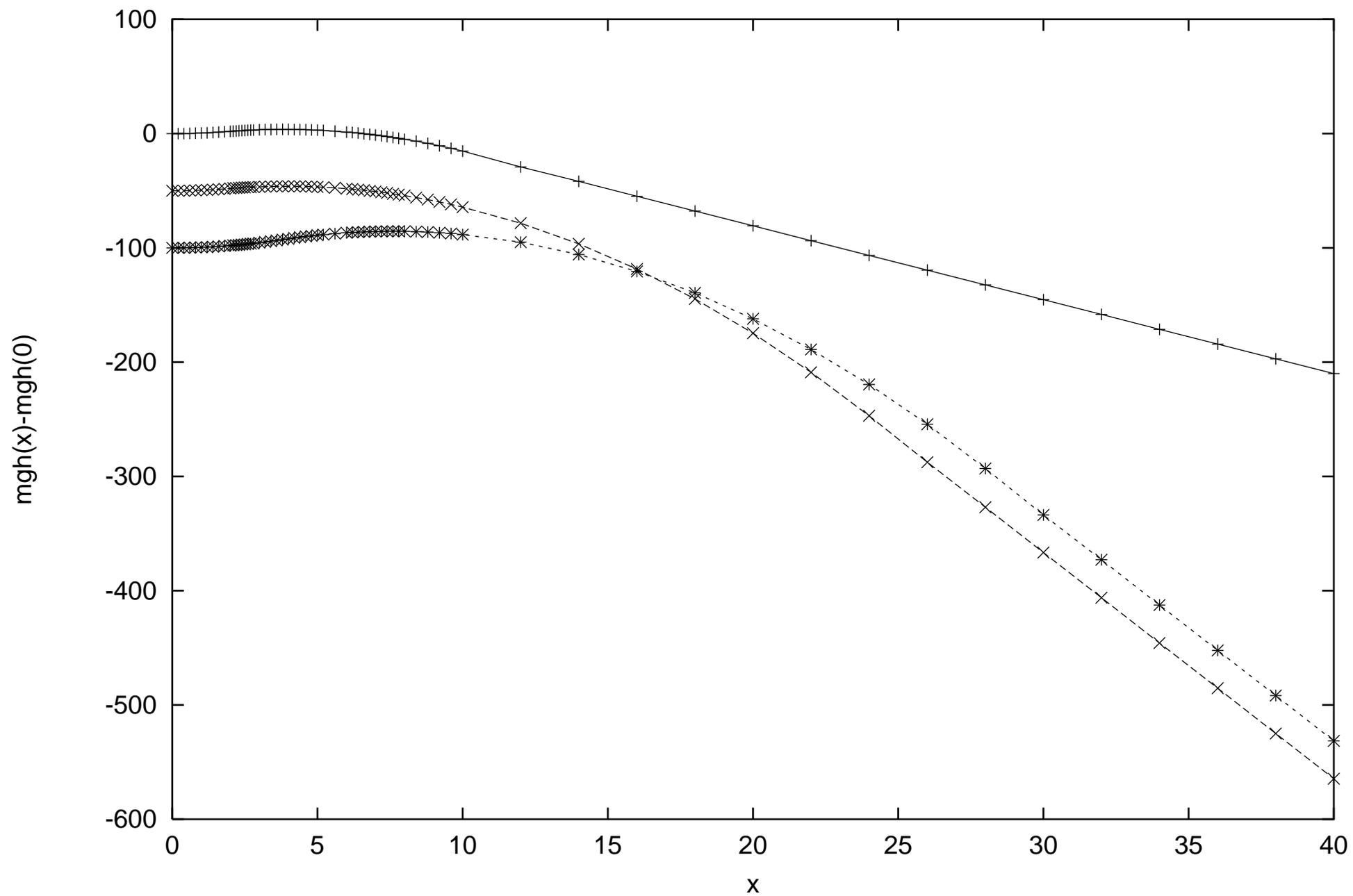
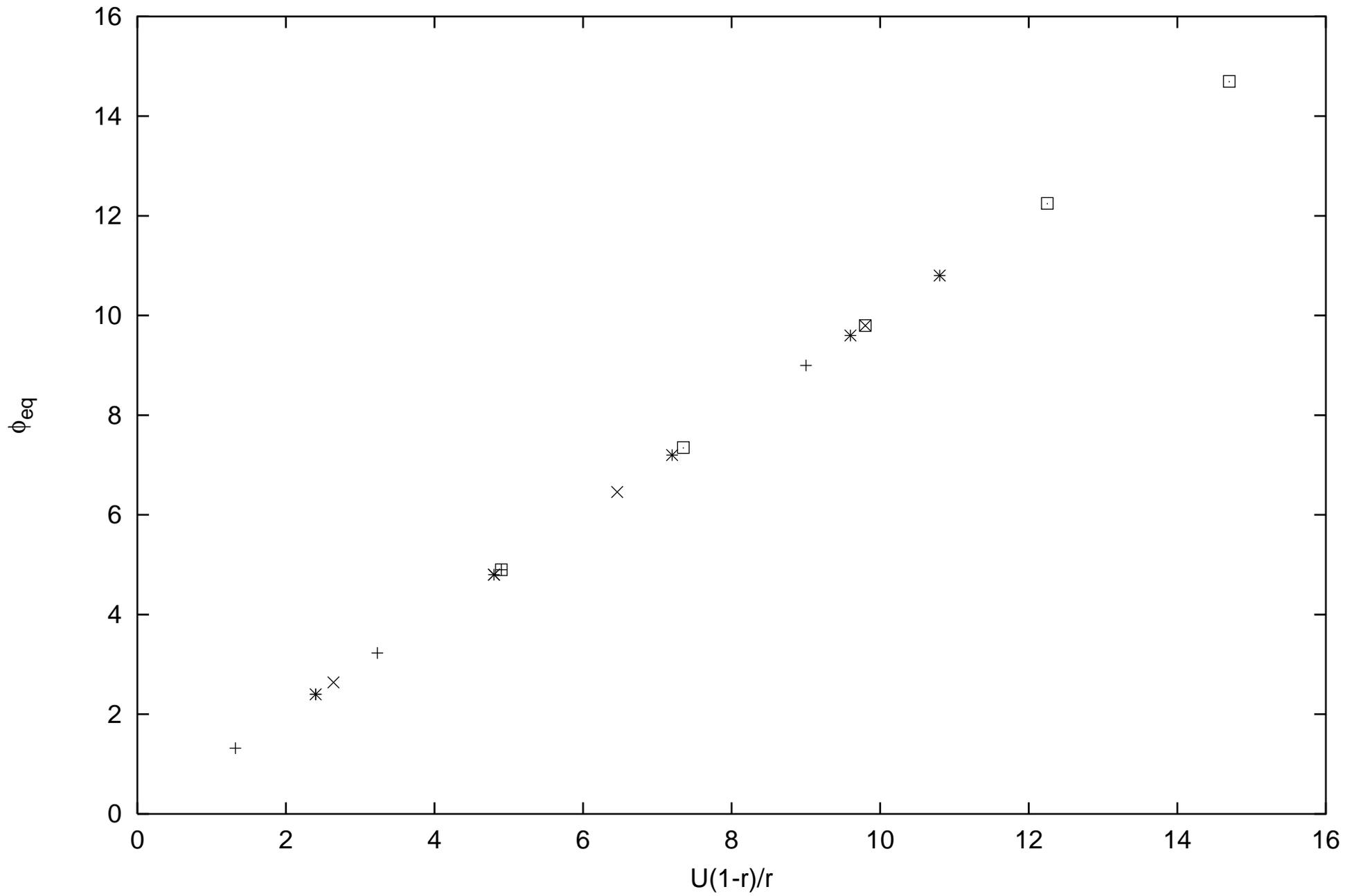
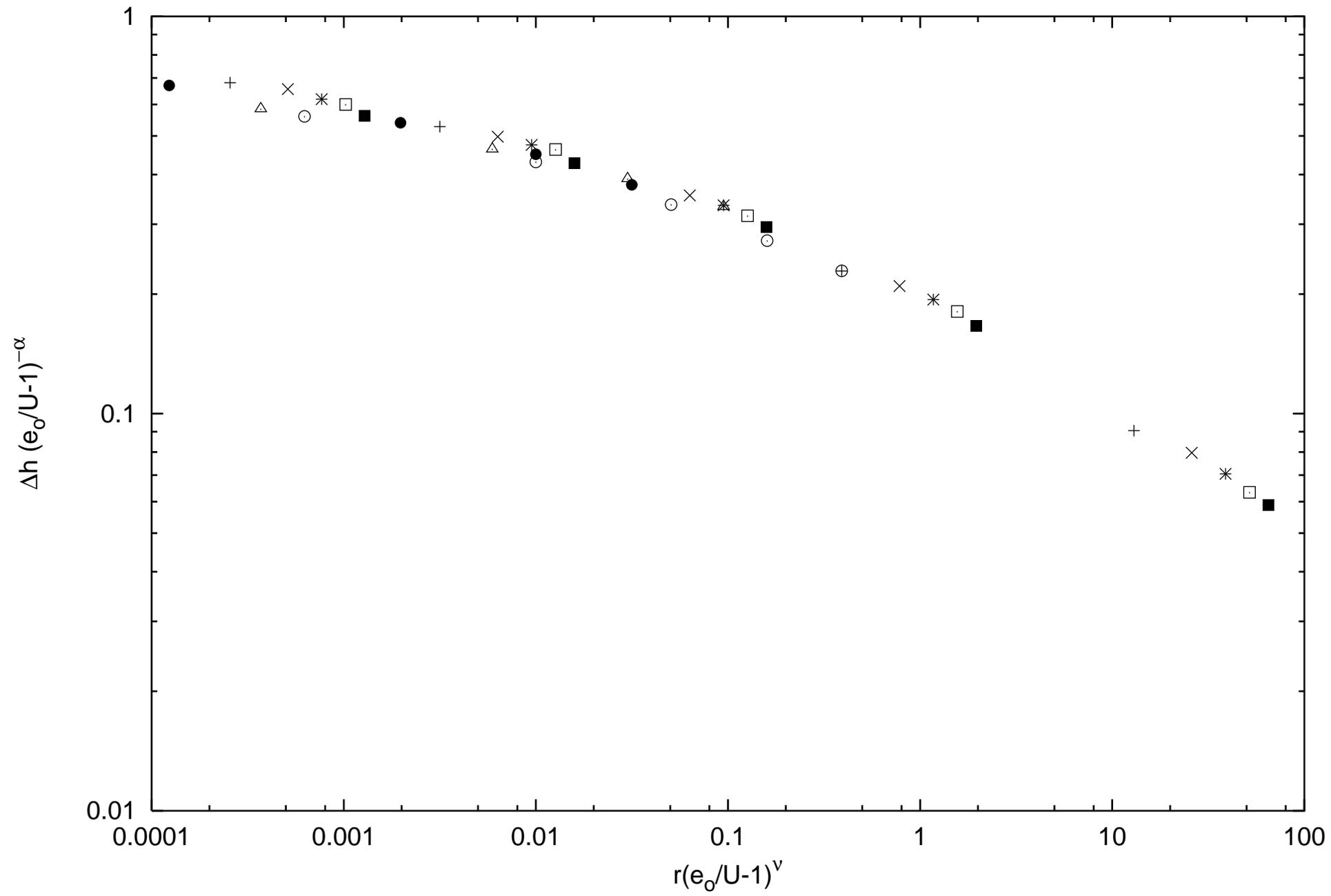
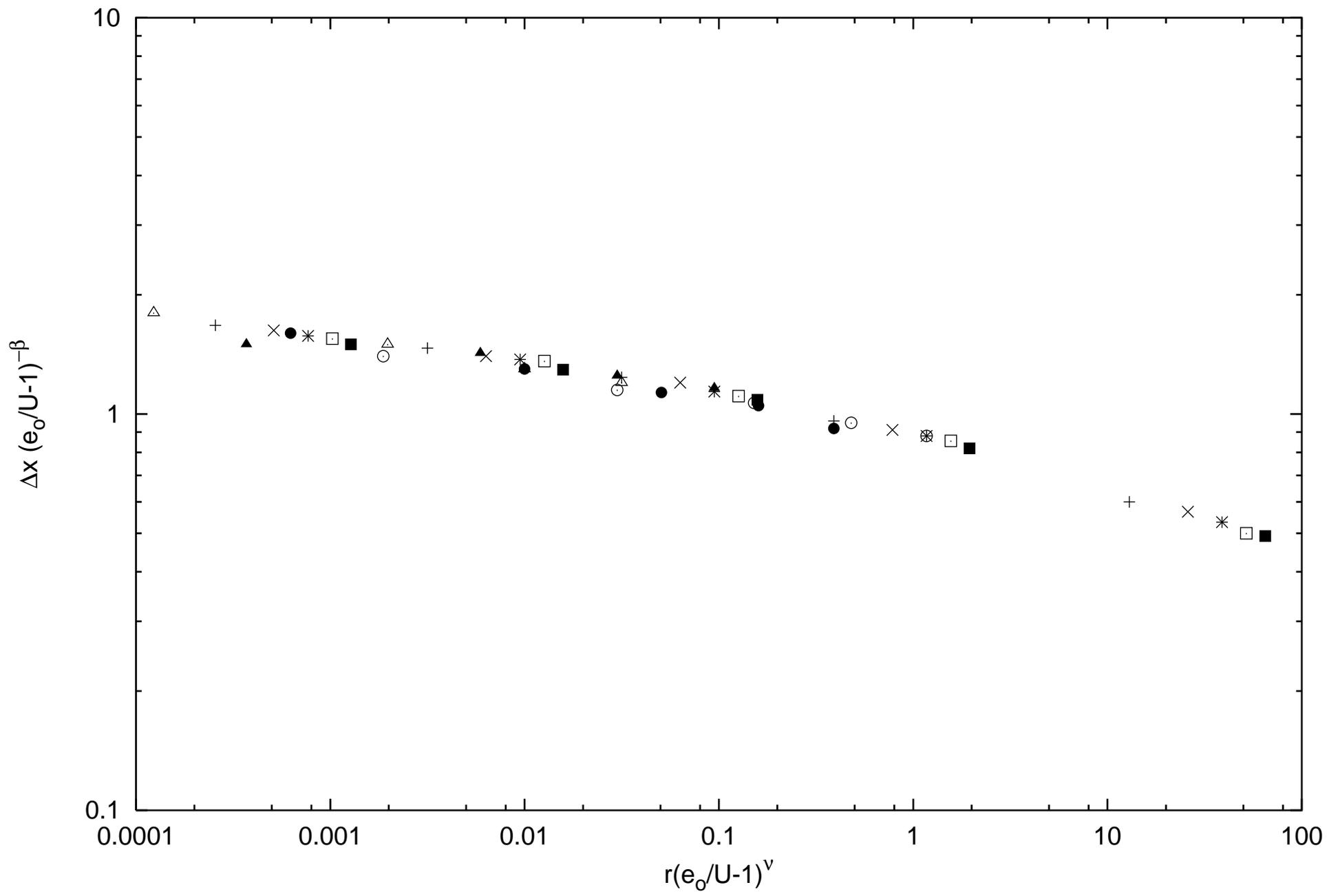
\end{document}